\documentstyle[12pt]{article}
\topmargin - 1.5cm
\leftmargin - 1cm
\begin{document}
   
\title{ Toward the Quantum Design of
Multichannel Systems \\
{\small The Inverse Problem Approach}}
\author{V.\ M.\ Chabanov, B.\ N.\ Zakhariev, and I.\ V.\ Amirkhanov} 
\date{}
\maketitle
\begin{center}
Laboratory of Theoretical Physics, \\
Joint Institute for Nuclear Research,  \\ 
141980 Dubna, Russia   \\
email: zakharev@thsun1.jinr.ru; URL http://thsun1.jinr.ru/$\sim$ zakharev/

\end{center}

\begin{abstract}
The multichannel generalization of the
theory of spectral, scattering and decay control is presented.
New  universal algorithms of construction of complex quantum systems 
with given properties are suggested. Particularly, transformations of 
interaction matrices leading to the concentration of waves in a chosen 
partial channel and  spatial localization are shown.  The limiting 
instructive cases illustrating different phenomena which ooccur with the 
combination of 'incompatible' properties  are considered.  For example, the 
scattering solutions with different resonance widths at the same energy for 
the same interaction are revealed.  Analogously, a 'paradoxical' 
coexistence of both strong reflection and absolute transparency is 
explained.  The case of the violation of 'natural' asymptotic behavior of 
partial wave function is demonstrated :  it has a greater damping decrement 
for the channel with a lower threshold.  Peculiarities of the multichannel 
periodic structures, bound states embedded into continuum, resonance 
tunneling and degeneracy of states are described.
\end{abstract}

\section{INTRODUCTION
    }
 
 Many years ago this journal published  Feshbach's paper "Unified theory of
nuclear reactions" [1] which attracted  attention  of a wide physical
community. This approach is a mighty tool for describing quantum systems
(e.g. few-body) with internal degrees of freedom.
 It can simplify the Schr\"odinger equation in partial derivatives reducing
it to a system of coupled {\it ordinary} differential equations
for the {\it vector-valued} wave functions, see (\ref{system}). 

The theory has gradually been developing in the direction of
its greater unification. Generalization to reactions with the
rearrangement of particles has been suggested by us in this journal [2-4].
A new contribution to the theory consists in employing the
inverse  problem (IP) and  quantum supersymmetry (SUSYQ) approach. We get 
a deeper insight into a relationship between interactions (${\bf V}$) and
observables ${\bf S}\equiv \{E_{n},\,C_{\alpha 
n}\}$:  spectral weights $C_{\alpha n} \enskip $ or 
$M_{\alpha n}$, energy levels and their analogs, e.g., resonance 
positions. 

In the IP and SUSYQ, the  {\it input}
data are ${\bf S}$. This is a remarkable advantage. It provides us with a 
{\it complete set} of elementary quantum transformations 
corresponding to variations of individual spectral parameters. 
They are given by explicit expressions: {\it the exactly solvable}
models, see, e.g., [5-7] and references therein for the one-channel case. 
For comparison, in the direct problem,  variation of {\it only one} chosen 
${\bf S}$-parameter is an extremely difficult problem.

There is in principle a possibility of converting any system
into another one of a different nature. Specifically, a 
scattering state can be 'raked up' into a bound state embedded into 
continuum (${\bf S}$-transformation at a single energy point without 
any other spectral  variation), see below.

A lot of qualitative physical information is revealed via a computer
visualization of these transformations. So, in the one-dimensional and
one-channel case we have developed a qualitative theory of quantum design
[8-12].
It becomes clear what "bricks and blocks" are needed to build systems
with given physical  properties. A comparison  with
a children's toy constructor-set suggests itself. In this way, the 
understanding of wave
mechanics is raised to the intuitive level. We acquire the possibility
of a qualitative foresight of  the results of IP-SUSYQ
transformations even without formulae and computer calculations. Namely, we
look for universal rules of the structural and dynamic response of 
quantum complexes when varying ${\bf S}$. Particularly, an arbitrary energy 
level $E_{n}$ can be  created, removed or shifted. Any object is uniquely 
 determined by a complete set of ${\bf S}$ data. So, it allows the choice of 
interactions providing the prescribed properties of the system.

The same status for the multichannel theory is still  to be achieved.
The one-channel IP {\it scalar} 
transformations represent a zero-measure  subset of their multichannel 
{\it matrix} analogs. Thus, one can expect  substantial  
advancement   and greater diversity of new effects in 
comparison with the one-channel case. 

 In this paper,  the latest 
results on the multichannel quantum design are discussed. 
First, we  give  
short preliminary information about the systems of coupled Schr\"odinger 
and IP equations.

\section{Multichannel equations of direct and inverse problems
}

The description of a wide class of quantum  objects with the multidimensional
wave functions  $\Psi $ can be
reduced to  the   system of close coupled Schr\"odinger (multichannel)
equations if we choose one variable $x$ and expand
$\Psi $ over the set of known basis functions $\Phi_{\alpha}(\xi )$ of
all other variables $\xi $
\begin{eqnarray}
\Psi (x,\xi )= \sum_{\alpha} \Psi_{\alpha}(x)\Phi_{\alpha}(\xi ).
\end{eqnarray}
The coefficients of these expansions, called channel wave functions, are
determined by the system of equations 
\begin{eqnarray}
-\Psi_{\alpha}''(x)+ \sum_{\beta}V_{\alpha\beta}(x)\Psi_{\beta}(x)
=E_{\alpha}\Psi_{\alpha}(x), \qquad E_{\alpha}=E-\epsilon_{\alpha},
\label{system}
\end{eqnarray}
\begin{eqnarray}
V_{\alpha \beta }(x)=
\int \Phi^{*}_{\alpha }(\xi ) V(x, \xi)\Phi_{\beta }(\xi ) d\xi 
\end{eqnarray}
where $V(x, \xi)$ is the potential for the corresponding multi-dimensional 
problem and $V_{\alpha \beta }(x)$ is the interaction matrix with 
non-diagonal
elements coupling  the equations. Greek indices $\alpha , \beta $
mean the channel numbers, i.e., partial equation numbers. 
Symbol $E$ stands for a total energy value and  $\epsilon_{\alpha}$ are
the threshold energies. The potential matrices   $V_{\alpha \beta }(x)$  
depend on $x$ locally.  

The IP gives the recipe of transformation from the initial 
$\stackrel{\circ}{\Psi }_{\alpha}(x)$, $\stackrel{\circ}{V}_{\alpha\beta}(x)$
to the new $\Psi_{\alpha}(x)$, $V_{\alpha\beta}(x)$ with the given spectral
data -- see, e.g., \cite{ChS} for Eqs. (\ref{psi})-(\ref{Q})
\begin{eqnarray}
\Psi_{\alpha} (x) = \stackrel{\circ}{\Psi }_{\alpha}(x) +
 \int\limits_{x}^{\infty } \sum _{\beta} K_{\alpha\beta}(x,x')
\stackrel{\circ}{\Psi }_{\beta}(x')dx';
\label{psi}
\end{eqnarray}
\begin{eqnarray}
V_{\alpha\beta}(x)= \stackrel{\circ}{V}_{\alpha\beta}(x) - 
2 \frac{d}{dx} K_{\alpha\beta}(x,x).
\label{V}
\end{eqnarray}
 Here, the transformation  matrix kernel $K$ is determined by the IP
equation
\begin{eqnarray}
  K_{\alpha \beta }(x,x') + Q_{\alpha \beta }(x,x') +
\int\limits_{x}^{\infty } \sum _{\gamma} K_{\alpha \gamma }(x,y)
Q_{\gamma \beta }(y,x')dy =0.
\label{IE}
\end{eqnarray}
The input spectral information is contained in its matrix valued kernel
\begin{eqnarray}
 Q_{\alpha \beta }(x,x')= \nonumber \\
\frac{1}{2\pi }\int\limits_{-\infty }^{\infty }
 k_{1}dk_{1}\sum_{\alpha' \beta' }\stackrel{\circ} {F}_{\alpha \beta' }
(x,\hat k) \{\hat{k}^{-1/2}[\hat {\stackrel{\circ}{S}}(\hat k)-
{\hat S}(\hat k)]\hat{k}^{-1/2} \}_{\beta'\alpha'}
\stackrel{\circ}{F}_{\beta\alpha'}(x',\hat k) \nonumber  \\
+ \sum_{n } \sum_{\alpha'\beta'}\stackrel{\circ}{F}_{\alpha\beta'}(x,E_{n})
M_{\alpha'n}M_{\beta'n}\stackrel{\circ}{F}_{\beta\alpha'}(x',E_{n})
\nonumber \\
 - \sum_{n}\sum_{\alpha'\beta'}\stackrel{\circ}{F}_{\alpha\beta'}(x,
\stackrel{\circ}{E}_{n}) \stackrel{\circ}{M}_{\alpha'n}
\stackrel{\circ}{M}_{\beta'n}\stackrel{\circ}{F}_{\beta\alpha'}
(x',\stackrel{\circ}{E}_{n}).
\label{Q}
\end{eqnarray}
In this expression the {\it partial spectral parameter} $M_{\alpha n}$
stands for the factor in the asymptotic behavior of the partial wave in the
$\alpha$th channel of the normalized bound state at the energy $E_{n}$:
$\Psi_{\alpha}(x,E_{n}) \to  M_{\alpha n}
\exp(-\sqrt{\epsilon_{\alpha}-E_{n}} \enskip x)$, as $x \to \infty$.
We use $F_{\alpha\beta}(x,E_{n})$ to designate the  matrix Jost solutions
of (\ref{system}) defined at the energies of bound states:
$F_{\alpha\beta}(x,E_{n}) \to
\delta _{\alpha \beta } \exp(-\sqrt{\epsilon_{\alpha }-E_{n}} \enskip x)$,
$ x \to \infty$. Here, the
first index means the channel number (i.e. partial equation number
in (\ref{system})), the second one  designates the type of a
boundary condition.
In (\ref{Q}) the scattering matrix is  $\hat S(\hat k)$ and
$F_{\alpha\beta}(x,\hat k)$ is the matrix Jost solution for the continuous
spectrum. The sums in $Q$ correspond to energy levels of
bound states and the integral refers to the continuous spectrum.
Symbol $\hat k$ is a diagonal momentum matrix
$k_{\alpha}\delta _{\alpha\beta}$,
$k_{\alpha}=\sqrt{E_{\alpha}}$. 

If  initial and  final systems differ in only a {\it finite} number of
discrete spectral parameters, {\it the kernel $Q$ becomes degenerate and
equation} (\ref{IE}) {\it reduces to  exactly solvable} algebraic equations
(see, e.g. \cite{AZ}).
Indeed, in this case the integral term in (\ref{Q}) vanishes because
$\stackrel{\circ}{S} \enskip = \enskip S$ and the terms in the sums with
$\stackrel{\circ}{E}_{n } \enskip = \enskip E_{n}; 
\stackrel{\circ}{M}_{\alpha n} \enskip = \enskip M_{\alpha n}$ are 
{\it cancelled}. So, there remain a {\it finite} number of terms with the 
{\it factorized} dependence on $x,x'$. Thus, IP gives rise to an infinite 
number of exactly solvable models which form a complete set, i.e. there is 
a possibility (at least theoretically) to fit any quantum system by them.

The above formulae correspond to the case when the spectral parameters
($S$, $M$) characterize the asymptotic ($x \to \infty$) behavior of the wave
functions.
This is Marchenko's approach (M). It is often convenient also to control the
spectral parameters which determine the wave behavior at the origin
according to Gelfand-Levitan's approach (GL). Then, instead of $M_{\alpha 
n}$ the derivatives of normalized bound states at $x=0$ are used 
$$C_{\alpha n}=\frac{d}{dx}\Psi _{\alpha}(x,E_{n})|_{x=0}. $$ The 
matrix-valued regular solutions  $\Phi _{\alpha \beta}(x,E_{n})$ enter in 
GL formalism instead of the Jost ones ($\Phi _{\alpha 
\beta}(x,E_{n})|_{x=0}=0$, $\frac{d}{dx}\Phi _{\alpha 
\beta}(x,E_{n})|_{x=0}=\delta _{\alpha \beta }$), and \begin{eqnarray} \Psi 
_{\alpha}(x,E_{n})= \sum_{\beta }\Phi _{\alpha \beta}(x,E_{n}) C_{\beta n}.  
\label{psgl} \end{eqnarray} Furthermore, the minus sign in (\ref{V}) is 
replaced by  the opposite one and the integration is performed over the 
interval $[0,x]$.  In what follows $C_{\beta n}$  will be referred to as a 
spectral weight vector (SWV) related to the nth eigenstate.

 For the reader interested in the IP approach it would be useful to
conceive the IP as the Gram-Schmidt orthogonalization
of the set of eigenvectors generalized to the case of infinite dimensions and
continuous basis. The bound state wave functions are considered as vectors 
in a special  Hilbert space in which the coordinate $x$ numbers the
vectors while the energy indices $n$ indicate their components.
The inner  product  in that space is determined by a measure
dependent on the SWV vectors $C$ or $M$ etc,  \cite{ZS}.
In fact,   Parseval's completeness relation
\begin{eqnarray}
\sum_{n}\Psi_{\alpha }(x,E_{n}) \Psi_{\beta }(x',E_{n})=\delta_{\alpha \beta}
\delta (x-x')
\label{par}
\end{eqnarray}
can be rewritten using (\ref{psgl}) as follows :
\begin{eqnarray}
\sum_{n}\sum_{\gamma \gamma' }\Phi _{\alpha \gamma}(x,E_{n}) C_{\gamma n}
C_{\gamma' n} \Phi _{\beta \gamma' }(x',E_{n}) =
\delta_{\alpha \beta } \delta (x-x').
\label{npar}
\end{eqnarray}
The last equality represents the orthogonality relation with the left
part as an 'inner product' for vectors $\Phi _{\alpha \gamma}(x,E_{n})$.
It is determined via summing over the multi-index ($\alpha ,n$)
with a weight $C_{\gamma n} C_{\gamma' n}$ which gives the measure of a 
new Hilbert space. The IP transformation $\stackrel\circ{C}_{\alpha n} \to 
C_{\alpha n}$  means the change of the measure. Then, the 
Hilbert space is changed so that the initial orthogonality  relation is 
violated. For this relation to be restored, the Gram-Schmidt 
orthogonalization is performed. So the IP equation (\ref{IE}) itself is 
just the equation for the coefficients of such a procedure.

 It turns out that
the exactly solvable models of the IP can be alternatively derived 
within the framework of the {\it supersymmetric quantum mechanics} 
(SUSYQ) formalism \cite{Witten} by E.Witten. The SUSYQ has been generalized 
to the 
multichannel case [15-19]. In many respects the SUSYQ  scheme is the 
same in both one- and multichannel cases.
The initial Hamiltonian, the second order differential operator,
is factorized into the simpler first order matrix operators $A^{\pm}= 
\pm \partial + W(x)$ ($\partial$ is a symbol of the derivative, $\enskip 
ImW(x)=0, \enskip \{W\}^{T}(x)= W(x) $) \cite{IH}. The  transformation
consists in permutation of $A^{\pm}$ :  \begin{eqnarray} H_{0} = A^{+} 
A^{-} + {\cal E} \rightarrow H_{1} = A^{-} A^{+} + {\cal E}, \end{eqnarray} 
where $\cal E$ is a constant factorization energy. 

The matrix $W(x)$ has the form as in the one-channel case \cite{Suk}
$$W(x)=\Psi_{0}'(x,{\cal E}) \Psi_{0}(x,{\cal E})^{-1}$$
where $\Psi_{0}(x,{\cal E})$ is an arbitrary matrix solution for the initial 
system at  the factorization energy. It should be noted here that the choice
of ${\cal E}$ and the selection of $\Psi_{0}(x,{\cal E})$ are crucial
points predetermining the resulting properties of the transformed Hamiltonian
and corresponding solutions.
It is remarkable that there is a simple relation between the solutions 
$\Psi_{0}(x,E)$ and $\Psi_{1}(x,E)$ of the Schr\"odinger equation with 
the initial and new Hamiltonians $H_{0},\,H_{1}$ \cite{Darboux}

\begin{eqnarray}
\Psi_{1}(x,E)=A^{-} \Psi_{0}(x,E)=(-\partial + W(x)) \Psi_{0}(x,E).
\end{eqnarray}

Let us perform the SUSYQ transformation  twice  at the same
energy ${\cal E}$ which can be arbitrary. The resultant formulae
can be equivalent to the IP ones for variation of  spectral weight or
for creation (removal) of any bound state while all other spectral 
parameters from the complete set $\{E_{n},C_{n}\}$ or $\{E_{n},M_{n}\}$ 
remain unchanged \cite{CZcc} (see also references for the one-channel case 
\cite{B}).  But there exist SUSYQ transformations which are 
non-equivalent to the IP ones. For example,  one can 
accomplish the double-SUSYQ procedure at different energies so that the 
first step is made at a bound state energy while the second one is 
performed at another energy, see, e.g., for the one-channel case the work 
\cite{PT}. Then a chosen bound state energy is shifted and the spectral 
weights for {\it all} the energy levels are changed. As infinite 
number of spectral parameters are involved, to fulfill such a 
transformation is infeasible in the 
framework of the IP.  Thus, in some cases, the SUSYQ formalism and 
one-parametric IP lead to the same results, whereas in some cases, SUSYQ 
transformations are not equivalent to finite-parametric models of the IP. 

The interaction matrices $V_{\alpha \beta }(x)$ can also correspond 
to non-local potentials $V(x,\xi ,\xi ')$ of the related
multi-dimensional problem. So,  wide classes of exactly 
solvable models mentioned above open new possibilities to
investigate systems with non-local forces which 
are still very poorly understood. 

\section{Uncoupled channels
}
   Let us start the consideration of the multichannel peculiarities  with
the limiting case of disconnected equations (\ref{system}) when $V_{\alpha
\ne \beta } \equiv 0$. 
The  partial channel transformed potentials and the corresponding wave
functions can be unified in the matrix, e.g., 2$\times$2 form 
\begin{eqnarray}
\hat V(x) = \left( \begin{array}{cc}
V_{11}(x)& 0\\
0 & V_{22}(x)
\end{array} \right),
\end{eqnarray}
\begin{eqnarray}
\hat \Psi(x) = \left( \begin{array}{cc}
\Psi_{1}(x)& 0\\
0 & \Psi_{2}(x)
\end{array} \right).
\end{eqnarray}
For uncoupled channels it is especially clearly seen that {\it the 
spectrum consists of several branches} (each for its channel).  By altering a
chosen threshold value $\epsilon_{\alpha }$, one can shift the $\alpha $th
partial branch with respect to the others. Some bound state energy
level of one spectral branch can occur in the continuous spectrum  of 
another channel. It is the simplest case of a bound state embedded 
into the continuum (BSEC).  In the general case of coupled channels, the 
spectrum becomes unified and its branches are not so clearly separated.

 It is interesting that one can join the initially independent branches at  
energy  $E_{n}$ through creating at this point a common bound state (or 
  BSEC if $E_{n}$ is above at least one of the thresholds).  This is attained in the IP formalism when both components 
$C_{\alpha n}$ of the SWV for $E=E_{n}$ are chosen to be non-zero. It 
results in some channel coupling $V_{12}(x)$ according to, e.g., Eqs.  
(36)-(38), section 4.7.  At other energies, spectral characteristics 
remain unchanged and correspond to initial separated channels despite 
arbitrarily strong coupling.  And only at the chosen energy $E_{n}$  the 
spectra are mixed completely.  For more details on the SWV control, see 
section 4.3.

The system of uncoupled channels is a convenient bridge linking
 one- and multichannel theories. Here the one-channel 'bricks and building
blocks' \cite{ChZ,CZ1,CZ3,CZ4,Z} appear in their pure form.

As the first example, we shall consider the two-channel model with
infinite rectangular potential wells $V_{11}(x)$, $V_{22}(x)$ having purely
discrete partial spectra.   The decrease in the  spectral
weight parameter $C_{11}$ for a bound state in the first channel is 
demonstrated in Fig.1.  This makes the chosen normalized partial wave 
 smaller at the origin and bigger on the right due to conservation of the 
 norm. It means shifting the chosen partial state in the configuration 
space (along x) to the right. The corresponding potential transformation 
$\Delta V_{11}(x)$ has a barrier on the left which pushes the wave to the  
right.  But this barrier alone would shift upwards all the energy levels of 
the first channel and violate the restriction that all other spectral 
parameters except for $C_{11}$  remain unchanged.  To compensate this 
undesirable influence of the barrier, the additional well is introduced on 
 the right. 

{\it A general rule-I for the spatial shift of an arbitrary partial state
in any potential} was found \cite{ZS,ChZ}. {\bf By decreasing (increasing)  
$C_{\alpha n}$ the $n$th partial state is shifted to the right (to the 
left) by n analogous $\Delta V_{\alpha \alpha}(x)$-blocks:  'barrier+well' 
('well+barrier') one for each bump of the wave function}.  Returning to the 
 initial system obviously requires the inversion of the perturbation  
block: the first transformation (barrier+well) combined with the second 
(the reverse 'well+barrier') cancel one another and restore the initial 
rectangular potentials.

 {\it The increase (decrease) in 
the spectral parameter $C_{11}$ at one
boundary of the allowed interval of motion is  equivalent to decrease
(increase) in the adjoint spectral parameter $M_{11}$ defined at another
boundary} (or for $x \to \infty $). The same is not valid
when an interchannel coupling is switched on (see below).

Shifting a chosen energy level over the $\alpha$th branch
of the spectrum is another group of the complete set of elementary
transformations. The case of potential symmetry with respect to the 
center of the interaction interval is especially simple \cite{ZS,ChZ,PT}.
Here, keeping the spectral weights unaltered is superfluous because 
nothing but energy levels  form a complete set of spectral parameters for
an  uncoupled channel.

{\it The qualitative rule-II}. {\bf Upward (downward) 
shifts of the energy level $E_{n}$ require barriers (wells) in the 
potential $\Delta V_{\alpha \alpha}(x)$ at the regions of the most 
 sensitivity of the $n$th chosen state: near the extremum of each 
anti-knot. Compensating wells (barriers) at positions of knots are 
necessary to keep other energy  levels at the same positions}. It is 
illustrated in Fig.1 for the lowest bound state in the second channel. 

Let us give a notion of analytic expressions for these transformations.
The formulae for the transformed potential and
the wave function of the lowest bound state related to the case
of changing $C_{11}$ are :
\begin{eqnarray}
V_{11}(x)=\stackrel{\circ}{V}_{11}(x)
-2\frac{d}{dx}\frac{(C_{11}^{2}/
\stackrel{\circ}{C}_{11}^{2}-1)
\stackrel{\circ}{\Psi }_{1}^{2}(x)}{1+
(C_{11}^{2}/\stackrel{\circ}{C}_{11}^{2}-1)\int_{0}^{x}
\stackrel{\circ}{\Psi }_{1}^{2}(y)dy},
\label{vaa}
\end{eqnarray}
and
\begin{eqnarray}
\Psi_{1}(x)=
\frac{(C_{11}^{2}/\stackrel{\circ}{C}_{11}^{2}-1)
\stackrel{\circ}{\Psi }_{1}(x)}{1+
(C_{11}^{2}/\stackrel{\circ}{C}_{11}^{2}-1)\int_{0}^{x}
\stackrel{\circ}{\Psi }_{1}^{2}(y)dy}.
\label{paa}
\end{eqnarray}
 It is remarkable that the new system is constructed of the initial 
$\stackrel{\circ}{\Psi }_{1}(x)$ and $\stackrel{\circ}{V}_{11}(x)$
as building material.
 Analogous simple formulae for the one-channel energy shift are given in 
\cite{ChZ,PT}.

 The same rule for spatial shifts is valid in the case when there exists 
continuous spectrum. If the external potential wall is of finite 
height, the last well of the $n$th block can be pressed out of the initial 
potential (carrier potential wells of the soliton-like form -- the 
fundamental quantum brick).  It is shown for the second channel in Fig.2.

In the limit $C_{\alpha n} \to 0$ or $M_{\alpha n} \to \infty$,
the bound state is carried away to infinity inside the reflectionless well. 
It is  equivalent to the effective {\it removal of the chosen energy 
level from the initial spectrum} \cite{ChZ}.

A {\it scattering state} wave can be considered as a bound state in the
infinite broad potential well. If we imagine it to be distributed over the 
infinite interval with norm 1, it has an infinite small 'spectral weight'
$\stackrel{\circ}{C}_{\alpha E} $.  By making  $C_{\alpha E}$ finite, 
this state {\it can be raked up to the origin and becomes  BSEC}. But 
there is an infinite number of bumps in the initial state. So we need an 
infinite set of the potential  well-barrier blocks, as shown in 
Fig.2. It is the limiting  case of the above-mentioned rule-I.  
  The BSEC  in one channel can coexist with a
scattering partial component in another channel at the same energy. It 
 is a non-trivial fact that this 
can be realized  for strong coupling of 
channels as well, see below.

It is remarkable that energy levels can also be shifted in the "imaginary
direction" \cite{CZ4}: $E \to E + i \Gamma $. The above stated 
qualitative rule is
valid  for $Im \Delta V_{\alpha \alpha }(x)$. So quadratically integrable 
decaying states
(with the exponential factor $\exp(-\Gamma t)$) are  new
convenient exact models. By analogy with the optical model it is 
an effective account of coupling with virtual channels not exhibited 
explicitly. 

\section{Coupled channels
}
As in the one-channel case, we can perform here a complete set of  
utmost elementary transformations.  Any chosen bound state energy level 
 can be shifted and  any partial channel
component $C_{\alpha n}$ of the SWV can be changed using the closed  
analytic expressions, see \cite{AZ,Z} and references therein. We shall 
consider some qualitative properties of multichannel solutions after some 
preliminary remarks.

Switching on non-diagonal elements of the interaction matrix in 
(\ref{system}) leads to a wave exchange between the channels which 
resembles a fluid motion in  communicating vessels. The conservation of a 
partial channel flux is usually violated and only the  total current is 
conserved.

Below are some simplest examples illustrating the specific 
influence of channel coupling. If we introduce a constant coupling 
$V_{12}=V_{21}=W \equiv const$ into two free equivalent 
equations of motion with initial solutions $\sim \sin(kx),\,\,  k=\sqrt{E}$ 
\begin{eqnarray} -\Psi_{1}''(x)+  W \Psi_{2}(x) =E\Psi_{1}(x) \nonumber \\ 
-\Psi_{2}''(x)+  W \Psi_{1}(x) =E\Psi_{2}(x),
\end{eqnarray}
we get new solutions with frequency that depends on the boundary 
conditions.  The partial waves  
coincide $\Psi_{1}(x)=\Psi_{2}(x)\, k=\sqrt{E-W} $
for identical partial boundary conditions. 
But if   the partial boundary conditions are 
chosen with  different signs the solutions $\Psi_{1}(x)=-\Psi_{2}(x) 
\sim \sin({\tilde k} x)$ have another  wave number 
${\tilde k}=\sqrt{E+W}$.

Let us consider a system on the whole line with equal thresholds and 
rectangular interaction matrix elements  $V_{11}(x)=V_{22}(x)= 
-V_{12}(x)=V, \enskip 0 \le x \le a$ ($V_{\alpha \beta }(x)=0, \enskip
x \le 0;\,\, a \le x$) 
\begin{eqnarray} -\Psi_{1}''(x)+ V 
\Psi_{1}(x) - V \Psi_{2}(x) = E \Psi_{1}(x) \nonumber \\ 
-\Psi_{2}''(x)+ V \Psi_{2}(x) - V \Psi_{1}(x) = E \Psi_{2}(x). 
\end{eqnarray} 
In this case, we have the absolute transparency  for identical boundary 
conditions in both the channels $\Psi_{1}(x)= \Psi_{2}(x) $. Moreover, 
there is free (!) motion due to complete cancellation of interactions. But 
if we choose different signs of the solutions $\Psi_{1}(x)=  -\Psi_{2}(x) 
$, there is no such cancellation, which results in {\it strong  reflection 
} of waves.

The next example demonstrates the {\it influence of channel
coupling on energy level spacing}.
Switching on the constant coupling $W$ between two channels with equivalent
thresholds and infinite rectangular potential wells $V_{11}=V_{22}$ results
in splitting initially degenerated levels $E_{n}=
\stackrel{\circ}{E}_{n} \pm W$.

Now let us consider the multichannel spectral control by using the
IP approach.
\subsection{Variation of SWV components 
}
Let us start with an arbitrary initial potential matrix 
$\stackrel{\circ}{V}_{\alpha\beta}(x)$ for the system with bound states.
We shall vary  SVW $\stackrel{\circ}{C}_{\alpha n} 
\rightarrow C_{\alpha n}$ 
and energy value $\stackrel{\circ}{E}_{n} \rightarrow E_{n}$ of the chosen 
$n$th bound state. The relevant expressions for the transformed potential 
matrix and solutions are derived, e.g., within the framework of the  
multichannel GL formalism. This is the case that corresponds to the exactly 
solvable models of IP discussed in section 2. Similar formulae are 
in the Marchenko approach as well.

The resulting potential matrix has the form
\begin{eqnarray}
V_{\alpha\beta}(x)= \stackrel{\circ}{V}_{\alpha\beta}(x) 
-2 \frac{d}{dx} \{\hat {\tilde 
\Upsilon}(x) \hat P(x)^{-1}  {\hat \Upsilon}(x)\}_{\alpha\beta},
\end{eqnarray}
where
\begin{eqnarray}
\hat {\tilde \Upsilon}(x)=\left( \begin{array}{cc}
\sum_{\beta} C_{\beta n} \stackrel{\circ}{\Phi}_{1 \beta}(x,E_{n})& 
\sum_{\beta} \stackrel{\circ}{C}_{\beta n} \stackrel{\circ}{\Phi}_{1 
\beta}(x,\stackrel{\circ}{E}_{n})\\
\sum_{\beta} C_{\beta n} \stackrel{\circ}{\Phi}_{2 \beta}(x,E_{n}) & 
\sum_{\beta} \stackrel{\circ}{C}_{\beta n} \stackrel{\circ}{\Phi }_{2 \beta}
(x,\stackrel{\circ}{E}_{n})
\end{array} \right),
\end{eqnarray}

\begin{eqnarray}
\hat \Upsilon(x)=\left( \begin{array}{cc}
\sum_{\beta} C_{\beta n} \stackrel{\circ}{\Phi}_{1 \beta}(x,E_{n})& 
\sum_{\beta} C_{\beta n} \stackrel{\circ}{\Phi}_{2 \beta}(x,E_{n})\\
-\sum_{\beta} \stackrel{\circ}{C}_{\beta n} \stackrel{\circ}{\Phi}_{1 \beta}
(x,\stackrel{\circ}{E}_{n}) & 
-\sum_{\beta} \stackrel{\circ}{C}_{\beta n} \stackrel{\circ}{\Phi }_{2 \beta}
(x,\stackrel{\circ}{E}_{n})
\end{array} \right),
\end{eqnarray}
and
\begin{eqnarray}
P_{11}(x)=
1+\int_{0}^{x} [\sum_{\beta} C_{\beta n} \stackrel{\circ}{\Phi}_{1 \beta}
(y,E_{n}) \sum_{\beta} C_{\beta n} \stackrel{\circ}{\Phi}_{1 \beta}(y,E_{n}) 
\nonumber \\
+ \sum_{\beta} C_{\beta n} \stackrel{\circ}{\Phi}_{2 \beta}(y,E_{n})
\sum_{\beta} C_{\beta n} \stackrel{\circ}{\Phi}_{2 \beta}(y,E_{n})]dy 
\nonumber \\
P_{12}(x)=-P_{21}(x) =
\int_{0}^{x} [\sum_{\beta} C_{\beta n} \stackrel{\circ}{\Phi}_{1 \beta}
(y,E_{n}) \sum_{\beta} \stackrel{\circ}{C}_{\beta n} 
\stackrel{\circ}{\Phi}_{1 \beta}(y,\stackrel{\circ}{E}_{n}) 
\nonumber \\
+ \sum_{\beta} C_{\beta n} \stackrel{\circ}{\Phi}_{2 \beta}(y,E_{n})
\sum_{\beta} \stackrel{\circ}{C}_{\beta n} \stackrel{\circ}{\Phi }_{2 \beta}
(y,\stackrel{\circ}{E}_{n})]dy  \nonumber \\
P_{22}(x)=1-\int_{0}^{x} [\sum_{\beta} \stackrel{\circ}{C}_{\beta n} 
\stackrel{\circ}{\Phi}_{1 \beta}(y,\stackrel{\circ}{E}_{n})
\sum_{\beta} \stackrel{\circ}{C}_{\beta n} \stackrel{\circ}{\Phi}_{1 \beta}
(y,\stackrel{\circ}{E}_{n})  \nonumber \\
+\sum_{\beta} \stackrel{\circ}{C}_{\beta n} 
\stackrel{\circ}{\Phi }_{2 \beta}(y,\stackrel{\circ}{E}_{n})
\sum_{\beta} \stackrel{\circ}{C}_{\beta n} \stackrel{\circ}{\Phi }_{2 \beta}
(y,\stackrel{\circ}{E}_{n})]dy.
\end{eqnarray}
The transformed regular solution at arbitrary energy E is
\begin{eqnarray}
\Phi_{\alpha\beta}(x,E) = \stackrel{\circ}{\Phi }_{\alpha\beta}(x,E)
-\int_{0}^{x}\sum_{\gamma} \{ \hat {\tilde \Upsilon}(x) \hat P(x)^{-1}  
{\hat \Upsilon}(y)\}_{\alpha\gamma} \stackrel{\circ}{\Phi }_{\gamma\beta}
(y,E)dy.
\end{eqnarray}
The solution $\Psi _{\alpha}(x,E_{n})=\sum_{\beta}C_{\beta n} 
\Phi_{\alpha\beta}(x,E_{n})=
\{\hat {\tilde \Upsilon}(x) \hat P(x)^{-1}\}_{\alpha1}$ represents 
a normalized wave function of the pushed nth bound state
$$\int_{0}^{\infty} \sum_{\alpha } [\Psi_{\alpha }(x,E_{n})]^2dx=1.$$
 Computer visualization of these formulae shows curious peculiarities
of the corresponding transformations of the potential matrix and 
wave functions some of which are presented here.

A gradual {\bf increase in only one partial channel component of 
the spectral weight vector $C_{\alpha n}$ or $M_{\alpha n}$  for the $n$th 
bound state results in progressive  concentration of waves in the 
$\alpha $th channel by their transition from all other partial channel 
components} $\beta \ne \alpha$ of the whole $n$th wave  function. So the 
waves are gathered from both the configurational and channel spaces while 
the total norm is conserved.  In the limiting case  $C_{\alpha n} 
\rightarrow \infty $ or $M_{\alpha n} \rightarrow \infty $, the whole wave 
function is pressed into the origin in the $\alpha $ channel or is shifted 
to infinity. It means that the chosen energy level effectively disappears 
from the spectrum. 
  A typical behavior of $V_{\alpha \beta }(x)$ and the wave 
function is shown in Fig.3  when $M_{11} \to \infty $.  The wave function 
is gradually moved away to  infinity by the reflectionless soliton-like 
'carrier' potential well. This process is accompanied by the new effect: the concentration 
of all waves in the chosen (first) channel at the expense of the others 
(second one).  This is true for both equal and different channel 
thresholds.  What is also amazing is that in spite of strong coupling of 
channels,  all other channels are almost completely emptied.  
  The 'carrier' interaction well is surprisingly similar to the 
one-channel case:  we meet again with {\it the soliton-like potential as a 
simple "brick" of quantum reconstruction}.  The discovery of this 
remarkable effect deepens our understanding of complex multichannel 
systems.  So in comparison with the one-channel case, there is a strong 
magnification of the response of the system to  increasing partial 
components $C_{\alpha n}$ or $M_{\alpha n}$.

 On the contrary, {\bf the gradual decreasing to zero of only one partial 
channel component of spectral  weight vector $C_{\alpha n}$ or
$M_{\alpha n}$ does not completely empty the 
$\alpha $th channel}, see Fig. 15 in \cite{CZ4} as a typical example of 
such a situation. The $\alpha $th channel wave is only partly shifted.
Some part of it is pushed out into other channels.
In the one-channel case, the $n$th state disappears if  $C_{n} \to 0$.  
In the multi-channel case, the zero value of the partial function
 and its derivative at one point does not mean that this wave must 
disappear. It returns to this channel at other points due to the wave 
exchange between channels.  Only if all the components  $C_{\alpha n}$ 
become zero, the bound state is completely removed.  
Thus, in comparison with the one-channel case, we have the weakening of the
response of the system to the decrease in the partial spectral weight 
parameter.  Furthermore, the equivalence of increasing 
(decreasing) in $C_{\alpha n}$ and  decreasing (increasing) in $M_{\alpha 
n}$ being valid in the one-channel case is violated when the channels are 
coupled.

The IP and SUSYQ  formalism allow one not only to remove energy levels
but also to create them at given positions. With the initial free motion
system this creation  leads to reflectionless interaction matrices.

\subsection{Transparency
}

It was once found that the Coulomb barrier between nuclei with large Z is
much more penetrable than was expected. Before our investigations
on  IP, we found the phenomena of intensified barrier
transparency and even supertransparency  in the multi-channel approach
to quantum many-body systems, see e.g. \cite{ZS}. 

Additional understanding  of the
phenomena can be achieved via consideration of the limiting model of 
total transparency. The isospectral transformation of a free
motion system  into the reflectionless one with a bound state gives 
\cite{ChZ2}
\begin{equation}
V_{\alpha\beta}(x)= 2 \frac{d}{dx} \frac{M_{\alpha} M_{\beta}
\exp[-(\kappa_{\alpha} + \nonumber \\
\kappa_{\beta}) \enskip x]}{1+\sum_{\gamma}
\frac{M_{\gamma}^{2}}{2 \kappa_{\gamma}} \exp(- 2 \kappa_{\gamma}
x)}, \qquad \kappa_{\alpha} = \sqrt{\epsilon_{\alpha}-E_{b}},
\label{VT}
\end{equation}
and
\begin{equation}
\Psi_{\alpha}(x,E_{b})= \frac{M_{\alpha} \exp(- \kappa_{\alpha} x)}
{1+\sum_{\gamma} \frac{M_{\gamma}^{2}}{2 \kappa_{\gamma}} \exp(-
2 \kappa_{\gamma} x)}.
\label{PT}
\end{equation}
Here $M_{\alpha }$ corresponds to the created bound state.
In \cite{ChZ2}, the transparent  interaction matrices in the case
of different thresholds were derived and shown. For equal thresholds, the
matrix elements of $V_{\alpha \alpha }(x)$ have a simple soliton-like form. 
For different thresholds, there occurs an  repulsion in 
$V_{\alpha \alpha }(x)$ which is unexpectedly necessary for the complete 
transparency. The waves reflected disappear via destructive interference  
with the backward waves decaying from other channels.  It is a 
non-trivial multichannel analog of the one-channel soliton-like potential 
well.

In (\ref{PT}),  the first channel component of the bound state wave 
functions decreases "unnaturally"  differently  in the directions $x \to 
\pm \infty $. It may seem that the channels become "disconnected" at large 
$|x|$ values, and partial waves must have a natural asymptotic  decrease 
$\exp(\mp \sqrt{E -\epsilon_{\alpha }}|x|),\,\, x \to \pm \infty $  for 
$V_{\alpha \beta }(x)$ rapidly decreasing in both the directions. But even 
 a weak coupling   $V_{\alpha \beta }(x) \to 0 $  as $x \to \infty $ can 
suck out the remainder waves from some channels into other ones violating 
the standard asymptotic behavior.  Instead of the "naturally expected" 
behavior \begin{eqnarray} \Psi_{\alpha}(x,E_{b}) \to \exp(- \kappa_{\alpha 
} x), \enskip x \to \infty \end{eqnarray} we have 
\begin{eqnarray}
\Psi_{1}(x,E_{b}) \to \exp[(- \kappa_{1} + 
2 \kappa_{2}) x], \enskip
 x \to -\infty
\end{eqnarray}
 and
\begin{eqnarray}
 \Psi_{2}(x,E_{b}) \to \exp( \kappa_{2} x), \enskip 
 x \to -\infty.
\end{eqnarray}
So,  there is a phenomenon of partial {\it inversion} of the "degree of
closeness" of different channels.

This can also be illustrated by reducing the two-channel system 
(\ref{system}) to one equation for the partial channel wave function 
$\Psi_{1}(x,E_{b})$. This is done by substituting the explicit expression 
for $\Psi_{2}(x,E_{b})$ in terms of $\Psi_{1}(x,E_{b})$ (see Eq. 
(\ref{PT})) \begin{eqnarray} \Psi_{2}(x,E_{b})= \frac{M_{2}}{M_{1}} 
\exp\{(- \kappa_{2} + \kappa_{1}) x\} \Psi_{1}(x,E_{b}).  \end{eqnarray} 
into the first equation
in (\ref{system}) with the potential matrix (\ref{VT}).
As a result, we get the {\it one-channel} Schr\"odinger equation for
$\Psi_{1}(x,E_{b})$ with the effective potential
\begin{eqnarray}
V(x)= 2 \frac{d}{dx}\{ \frac{M_{1}^2
\exp[-2 (\kappa_{1}) \enskip x]}{1+\sum_{\gamma}
\frac{M_{\gamma}^{2}}{2 \kappa_{\gamma}} 
\exp(- 2 \kappa_{\gamma} x)}\} + \nonumber \\
2 \frac{d}{dx}\{\frac{M_{1} M_{2}
  \exp[-(\kappa_{1}+\kappa_{2}) \enskip x]}
{1+\sum_{\gamma}\frac{M_{\gamma}^{2}}{2 \kappa_{\gamma}}
\exp(- 2 \kappa_{\gamma}x)}\} \exp[(- \kappa_{2} + \kappa_{1}) x]
\frac{M_{2}}{M_{1}}.
\label{efv}
\end{eqnarray}
As $x \to -\infty$, it becomes a non-zero constant  $4 \kappa_{2} 
(\kappa_{2} - \kappa_{1}) > 0$ whereas $V_{22}(x)$ decreases as 
$\exp[(2 \kappa_{2}-2 \kappa_{1}) x]$, which leads to the unnatural 
inversion of closeness of channels (on the left). This illustrates 
the fact that the rate of the asymptotic decrease in the partial channel 
wave function can be strengthened even by exponentially decreasing coupling 
$V_{\alpha \beta }$. The interaction (\ref{efv}) is {\it energy dependent}.

In \cite{CZcc}, we have also  found transparent interaction matrices 
without bound  states by employing the multichannel SUSYQ formalism which 
gives a broader class of exactly solvable models. {\it These quantum 
systems not previously known have no one-channel analogs}. They play 
a significant role in creating a bound state energy level close to the 
existing one in the initial spectrum (see the next section and the left 
part of Fig.5).

One of the multichannel peculiarities found by us is the  right-left symmetry
violation \cite{ZS,Z} for the reflection of a complex particle by a 
non-symmetrical potential barrier. It is more penetrable in one direction 
for an incident particle in the ground state. And in the opposite 
direction (in two-channel approximation) it is more transparent in the 
exited state.

Let us now  consider the model demonstrating shifts of  energy  levels
up to their degeneracy.
\subsection{Degeneracy of energy levels 
    }

In the one channel case, two bound states are forbidden to have the same 
energy. We have discovered  \cite{BDS} the 
phenomenon of "effective annihilation" of degenerating states.  
These states must both be equal, as solutions of 
the same equation with the same boundary conditions, and simultaneously 
 orthogonal as different states. This contradiction 
is resolved in the following way. There is a separation of parts of the 
waves by the non-penetrable barrier (infinitely high or broad) under 
which $\Psi_{\alpha } \to 0$. So the wave functions become identical up to 
the sign  in the limit of zero distance between the levels $E_{n}-E_{n+1} 
\to 0$.  But in the case of $M$ channels, $M$ degenerate  states can 
coexist provided that their SWV are  linearly independent. 

 Let us use the IP  formulae for 
creation of two energy levels $E_{b}$ and $E_{b}'$ which are very close 
to one another (including  the case of degeneracy). As  an
initial two channel system on the whole line we choose the free motion with 
$\stackrel\circ{V}_{\alpha \beta }(x)=0$.  These formulae are 
similar to those in  section 4.1:  
\begin{eqnarray} V_{\alpha\beta}(x) = 2 
\frac{d}{dx} \{\hat {\tilde \Upsilon}(x) \hat P(x)^{-1}  {\hat 
\Upsilon}(x)\}_{\alpha\beta}, \end{eqnarray} where \begin{eqnarray} \hat 
{\tilde \Upsilon}(x)=\left( \begin{array}{cc} M_{1} \exp(-\kappa_{1} x)& 
M_{1}' \exp(-\kappa_{1}' x)\\ M_{2} \exp(-\kappa_{2} x) & M_{2}' 
\exp(-\kappa_{2}' x) \end{array} \right), \end{eqnarray}

\begin{eqnarray}
\hat \Upsilon(x)=\left( \begin{array}{cc}
M_{1} \exp(-\kappa_{1} x)& M_{2} \exp(-\kappa_{2} x) \\
M_{1}' \exp(-\kappa_{1}' x) & M_{2}' \exp(-\kappa_{2}' x)
\end{array} \right),
\end{eqnarray}
and
\begin{eqnarray}
P_{11}(x)=
1+\sum_{\alpha} \frac{M_{\alpha}^{2}}{2 \kappa_{\alpha}} 
\exp(-2 \kappa_{\alpha} x) \nonumber \\
P_{12}(x) = P_{21}(x) = \sum_{\alpha} \frac{M_{\alpha} M_{\alpha}'}
{\kappa_{\alpha} \kappa_{\alpha}'} \exp[-(\kappa_{\alpha} + 
\kappa_{\alpha}') x] \nonumber \\
P_{22}(x)=1 + \sum_{\alpha } \frac{(M_{\alpha}')^{2}}{2 \kappa_{\alpha}'} 
\exp(-2 \kappa_{\alpha}' x).
\end{eqnarray}
Here $\kappa_{\alpha} = \sqrt{\epsilon_{\alpha}-E_{b}}$, 
$\kappa_{\alpha}' = \sqrt{\epsilon_{\alpha}-E_{b}'}$ and 
$M_{\alpha}$, $M_{\alpha}'$ are the SWV of the states under creation 
with the energies $E_{b}$ and $E_{b}'$, respectively.
Let us combine normalized vector-column wave functions at the energies 
$E_{b}$ and $E_{b}'$ to form the matrix $\hat \Psi(x)$ for 
compactification of the notation
\begin{eqnarray}
\hat \Psi(x) = \hat {\tilde \Upsilon}(x) \hat P(x)^{-1}.
\end{eqnarray}
When two levels $E_{b}$ and $E_{b}'$ with linearly dependent SWV  come 
closer and closer to each  other, the features of the effective 
annihilation become apparent. The wave functions are split and removed 
 (either their parts or the whole states), as in the one-channel 
annihilation case.  The same happens if  SWV of the degenerated states 
become more linearly dependent (see Fig.4). Rudiments of spatial separation 
(preparation to the effective annihilation) can be observed even without 
 changing the energy distance between the levels.  As in the one-channel 
case (\cite{CZ4}, Fig. 13), the states with linearly dependent SWV even 
corresponding to different energy levels will resist the concentration in 
the narrow spatial region.  There was found an exchange of knots between 
partial channel wave functions with  flips of some of their bumps 
(\cite{ChZ}, Fig.14).
The annihilation  phenomenon and the recoil of all 
states with quantum number $n \ne m$ from the origin when the chosen mth 
state is concentrated to $x=0$ \cite{ChZ,Z} have much in common. They
are connected with the resistance of system of different states to their 
concentration in phase space. 

There also occurs 'splitting' of BSEC  when $E_{1 BSEC} \to E_{2 BSEC}$ (as 
has been shown in \cite{CZ4}, Fig. 22).

Unexpectedly, when two levels even with 
{\it linearly independent} SWV come close to each other, this is often 
followed by  special 
evolution of the interaction matrix $V_{\alpha \beta }(x)$. In the case of 
unequal thresholds, there occurs a spatial separation of some block of 
$V_{\alpha \beta }(x)$, see  Fig.5.
This block is transparent and goes to infinity in the limit of the level
degeneracy resembling one-channel annihilation \cite{BDS}.
This separated block, in contrast with the annihilation case, does not 
carry away any part of the bound states. All bound states remain completely 
inside the main part of  $V_{\alpha \beta }(x)$. We have already mentioned 
this peculiar interaction  matrix in the previous section.
An explanation of this new effect is still an open problem.

It is interesting that, in the multichannel case, there is the
possibility of moving an energy level through the other level positions 
without the appearance of singularities in the interaction matrix.  Such an 
 operation is feasible if the SWV's of the shifted state and the crossed 
 states are linearly independent. The crossed states must be less then 
M-fold degenerate where M denotes the total number of channels.

\subsection{Bound states embedded into the continuum 
}

BSEC being a wonderful phenomenon in the one-channel  case, 
has a diversity of new multichannel aspects.  A 
possibility appears to control the asymptotic decrease of BSEC:  making 
BSEC short- or long-range.  The examples of exponential BSEC's falloff 
$V_{\alpha \beta }(x)$ and $\Psi _{\alpha }(x,E_{BSEC}) $ for systems with 
BSEC below the threshold of the most closed channel were shown in 
 \cite{ZS}.  Later, we have discovered that in special cases BSEC  can 
decrease like $\sim 1/x$ \cite{bal} when its energy is between the 
threshold ones.  It can be explained as follows.  The  regular 
solutions increase exponentially in closed channels as $x \to \infty $. 
 The physical scattering solutions $\Psi_{\alpha }(x,E)$ are constructed 
as a linear combination of the columns of the matrix of regular solutions 
with the coefficients providing  exponential decrease in the closed 
channels. Let us choose the SWV components for BSEC proportional to 
these coefficients for solutions in the initial system (with non-vanishing 
channel coupling) where BSEC is to be created at $E_{BSEC}$. Then the 
resultant $V_{\alpha \beta }(x)$ and $\Psi _{\alpha}(x,E_{BSEC})$ have 
$\sim 1/x$ behavior.  It is caused by the {\it linear} increase of the 
integral $$\sum_{\alpha}\int_0^x{[\sum_{\beta}C_{\beta}\stackrel{\circ } 
{\Phi }_{\alpha\beta}(y,E_{BSEC})]}^{2}\,dy$$ in the denominator of the 
formulae for $V_{\alpha \beta }(x)$ and $\Psi _{\alpha}(x,E_{BSEC})$ which 
are direct GL analogs of Eqs. (36) - (38), section 4.7. In this expression 
exponentially growing terms are cancelled by the aforementioned choice of 
SWV components $C_{\alpha }$. Any violation of the ratio between $C_{\alpha 
}$ leads to the exponential growth in the denominator.  This results in 
exponential falloff of $V_{\alpha \beta }(x)$ and $\Psi 
_{\alpha}(x,E_{BSEC})$.

 As in the one-channel case \cite{Sof}, a BSEC solutions above the 
thresholds, despite the different frequencies of oscillations of partial
wave components, have strict conformity between knots of functions 
and diagonal interaction matrix elements. For each partial function bump 
there is a relevant potential well-barrier block, see Fig.1 and \cite{Sof}.

We have considered in section 4.1  pumping waves 
into a chosen channel by increasing the corresponding component of SWV. 
Analogous effect can also be observed for BSEC.
  The BSEC can be obtained even by taking a single  partial 
$\alpha $th component of SWV to be non-zero. Then the interchannel coupling
provides the transformation of other partial scattering waves into 
BSEC.

\subsection{Phenomena of coexistence of states with 'incompatible'
properties
}
The existence of several physically allowed linearly independent solutions  
of Eq.  (\ref{system})  make it possible, unlike 
the one-channel case, to combine at the same energy different peculiarities 
of wave dynamics. For example, in \cite{CZcc} it was revealed that 
a bound state can coexist with a scattering state even 
without a strict space separation of propagating and bound waves.  The 
qualitative arguments to resolve this "paradox" are the following.  In the 
M-channel  case, there  exist M linearly independent regular solutions of 
(\ref{system}) for the same energy point.  Above all thresholds these 
solutions are scattering states. In the IP and SUSY approach one of the 
scattering states can be transformed into a BSEC at $E=E_{b}>\epsilon _{i}; 
\, i=1,2,...M$ ). Then there will remain $M-1$ scattering state solutions. 
As a result, {\it  it is impossible to 
construct scattering solutions at BSEC energy with arbitrary asymptotic 
conditions}, e.g., with the incident wave in one channel only.
 
By analogy with the effect considered above, {\it there can exist scattering 
solutions with  different resonance decay widths at the same energy}.  
We can construct such an exactly solvable model. Let us start from two 
uncoupled channels with different thresholds and resonances 
at the same energy with unequal widths $\Gamma _{1}, \Gamma _{2}$. We can
create a bound state  with non-zero SWV components 
in each channel  by using  the GL analogs of Eqs.(36)-(38). 
There arises a strong coupling $V_{12}(x)$ between channels. 
 For incident waves in the  $\alpha $th channel only, we shall get resonance 
scattering with the width $\Gamma _{\alpha }$ {\it despite the intensive 
exchange} of waves between the channels inside the interaction  region.  
Outgoing waves survive in the entrance channel only. After the mixing of 
channel waves in  region where $V_{12}(x) \ne 0$, they return to  
the entrance channel. It is a common opinion that  compound states 
(resonances) in complex systems "forget" how they were generated. 
Here an extremely simple example has been demonstrated of direct dependence 
of the life-time of a quasi-bound state on boundary conditions.  So in the 
general case the analogous phenomenon is also possible.

In section 4.7, the coexistence of transparency and strong reflection is 
shown.

\subsection{ M+1 spectra      
}

The systems with the pure discrete spectrum in the one-channel case are 
uniquely determined
by the set of $\{E_{n}, C_{n}\}$ or by {\it two spectra}  $\{E_{n}, 
E_{m}\}$. Here the index $m$ is related to eigenvalues with different 
boundary conditions at one end point of an interval (e.g., at the origin 
$\Psi'_{\alpha }(0)=0$ instead of $\Psi_{\alpha }(0)=0$) \cite{ChS}, p.401.  
Daskalov \cite{D} has suggested exactly solvable models corresponding to 
variations of any eigenvalues from two  spectra $\{E_{n}, 
E_{m}\}$.  This enlarges the possibilities to control quantum 
systems.  In the M-channel case,  the set of eigenvalues and the sets of 
$M$ SWV components $\{E_{n},C_{\alpha n}\}$ form the complete $(M+1)$-fold 
set of parameters. They are equivalent to  M+1 spectra with different 
boundary conditions \cite{ChZ}. Exact models for M+1 spectra variations 
are now under construction.  

\subsection{Resonance tunneling
}
Almost non-penetrable barriers could be combined in a system transparent 
at discrete energy values, see e.g. \cite{ZS,BG}.
 This phenomenon in the one-channel case has already found numerous 
applications, see  \cite{RT} and references therein. 
But we have not seen any papers on multichannel resonance tunneling.
The waves accumulated between the barriers in the $V_{\alpha \beta }(x)$ 
decay in both directions. The  decaying waves going backward can have the 
opposite phases and the same amplitudes as the reflected waves in all open 
channels. In this case reflected and decaying waves cancel one another 
totally, which results in resonance tunneling.  It appears that the 
multichannel analog of this phenomenon is restricted by additional 
conditions.  We have found that it can be achieved {\it only for special 
proportions} of amplitudes of incident waves in different 
channels at discrete resonance energies $E_{res}$. 

  So, there can 
coexist at the same energy solutions without reflection at all and with 
strong reflection depending on boundary conditions.  The relevant 
example (exactly solvable IP model) can be constructed of the two initially 
uncoupled channels ($\stackrel{\circ}{V}_{12}(x)\, \equiv \, 0$). We can 
choose the potential $\stackrel{\circ}{V}_{11}(x)$ in the first channel  
having "one-channel" resonance tunneling at $E=E_{res}$ and  in the 
second channel the potential $\stackrel{\circ}{V}_{22}(x)$ being weakly 
penetrable at the same energy.  Let us now create  a  bound state at 
$E=E_{b}<0$ common for two channels by the IP transformation 
$\stackrel{\circ}{V}_{\alpha \beta }(x)\rightarrow \, V_{\alpha \beta}(x)$ 
that mixes intensively partial channel waves (i.e., leads to $V_{12}(x) 
\enskip \ne \enskip 0$). Then the non-trivial two-channel system occurs 
for which the waves incident in the first channel (i.e., $\psi_{\alpha}(x) 
\rightarrow \delta_{1 \alpha} \exp(- i k_{1} x)$, $x \rightarrow \infty$) 
propagate without reflection at $E=E_{res}$ whereas the waves incident in 
the second channel (i.e., $\psi_{\alpha}(x) \rightarrow \delta_{2 \alpha} 
\exp(- i k_{2} x)$, $x \rightarrow \infty$) are strongly reflected.  This 
is possible because the IP transformations keep the continuous spectrum 
characteristics unchanged. The 
corresponding analytic expressions have the form :  
\begin{equation} V_{\alpha\beta}(x)= \stackrel{\circ}{V}_{\alpha \beta }(x) 
+ 2 \frac{d}{dx} \frac{\sum_{\alpha'\beta'} 
\stackrel{\circ}{F}_{\alpha\beta'}(x,E_{b}) M_{\alpha'} 
M_{\beta'}\stackrel{\circ}{F}_{\beta\alpha'}(x,E_{b})}{1+ \int_{x}^{\infty} 
\sum_{\gamma} \sum_{\alpha'\beta'} 
\stackrel{\circ}{F}_{\gamma\beta'}(y,E_{b}) M_{\alpha'}
M_{\beta'}\stackrel{\circ}{F}_{\gamma\alpha'}(y,E_{b})dy},
\end{equation}
where $\kappa_{\alpha} = \sqrt{\epsilon_{\alpha}-E_{b}}$;
\begin{eqnarray}
F_{\alpha\beta} (x,E) = \stackrel{\circ}{F}_{\alpha\beta}(x,E) - \nonumber 
\\ \int\limits_{x}^{\infty } \sum _{\gamma'} \frac{\sum_{\alpha'\beta'} 
\stackrel{\circ}{F}_{\alpha\beta'}(x,E_{b}) M_{\alpha'}
M_{\beta'}\stackrel{\circ}{F}_{\gamma'\alpha'}(x',E_{b})}{1+
\int_{x}^{\infty} \sum_{\gamma} \sum_{\alpha'\beta'}
\stackrel{\circ}{F}_{\gamma\beta'}(y,E_{b}) M_{\alpha'}
M_{\beta'}\stackrel{\circ}{F}_{\gamma\alpha'}(y,E_{b})dy}
\stackrel{\circ}{F}_{\gamma'\beta}(x',E)dx',
\end{eqnarray}
where $F_{\alpha\beta}(x,E)$ is the Jost solution at arbitrary energy E
obeying the asymptotic condition $F_{\alpha\beta}(x,E) \to \delta _{\alpha 
\beta } \exp(- i \sqrt{E- \epsilon_{\alpha }} \enskip x)$ as $ x \to \infty$.
The  created bound state has the form
\begin{eqnarray}
\Psi_{\alpha } (x,E_{b}) =  \frac{\sum_{\beta} M_{\beta}
\stackrel{\circ}{F}_{\alpha \beta}(x,E_{b})}{1+
\int_{x}^{\infty} \sum_{\gamma} \sum_{\alpha'\beta'}
\stackrel{\circ}{F}_{\gamma\beta'}(y,E_{b}) M_{\alpha'}
M_{\beta'}\stackrel{\circ}{F}_{\gamma\alpha'}(y,E_{b})dy}.
\end{eqnarray}

We have found in \cite{pen} that two coupled particles  may exhibit
resonance tunneling properties while undergoing transition through only one 
potential barrier. This problem can be reduced to multichannel equations 
with two barriers in the interaction matrix. So, our consideration above 
applies in this case as well.

\subsection{Periodic interaction matrices
}

As a particular limiting case of resonance multichannel tunnelling one can  
consider the wave motion in periodic fields. In this case the discrete 
resonance tunneling points for two-barrier system broaden into allowed 
continuous spectrum bands. 

To develop the theory of excitations of crystals, there  can be
useful exact solutions for quasi-particles with the inner (multichannel) 
degrees  of freedom.

Let us consider the two-channel system.
For two uncoupled channels with periodic $V_{\alpha \alpha }(x)$ there are
two independent spectral branches with energy bands of allowed and forbidden
motion with quasi-momenta $K_{1,2}$. For example, in the case of 
Dirac's comb potentials $V_{\alpha\alpha}(x)= 
\sum_{n=-\infty}^{\infty}V_{\alpha } \delta (x + n a)$,  each of the 
channels we have
\begin{eqnarray}
\cos(K_{\alpha } a) = \frac{\sin(k_{\alpha } a) V_{\alpha }}{2 k_{\alpha }} 
+ \cos(k_{\alpha } a),  \enskip  k_{\alpha}=\sqrt{E-\epsilon_{\alpha }}.
\label{comb}
\end{eqnarray}
One could suspect that switching on a weak coupling would hinder wave
propagation at the energies belonging to a forbidden zone in one channel 
and to an allowed zone in another one. Indeed, it may seem that the 
exponential increase in the wave function component in the forbidden zone 
would prevail and cause physically unacceptable asymptotical divergence of
solutions. But really it appears that switching on the coupling changes 
slightly the allowed zones of the uncoupled channels.  So just the regime of 
oscillations,  characteristic to the allowed zones, dominates. It 
results in damping the exponential increase corresponding to forbidden 
zones.  In fact, there are two types of two-channel generalized Bloch's 
waves 
\begin{eqnarray} \Psi_{B \alpha }^{(1,2)}(x) = \exp(\pm i {\cal 
K}^{(1,2)} a) \Psi_{B \alpha }^{(1,2)}(x-a) \label{bloch} \end{eqnarray} 
with the quasi-momenta ${\cal K}^{(1,2)}(E)$ common for both the channels.  
We can find such quasi-momenta by using, e.g., the exact model with the 
following periodic $\delta $-interaction matrix :  \begin{eqnarray} 
V_{\alpha \beta}(x) = \sum_{n=-\infty}^{\infty}V_{\alpha \beta} \delta (x + 
n a).  \label{vcomb} \end{eqnarray} With such a choice of $V_{\alpha 
\beta}(x)$ we have \begin{eqnarray} \cos({\cal K}^{(1,2)} a) = \frac{1}{2} 
\{\cos(K_{1} a) + \cos(K_{2} a) \nonumber \\ \pm \sqrt{[\cos(K_{1} a) - 
\cos(K_{2} a)]^2 + \frac{W^2 \sin(k_{1} a) \sin(k_{2} a)}{k_{1} k_{2}}  
}\}, \label{cocomb} \end{eqnarray} where $W 
\equiv V_{12}=V_{21}$, $V_{\alpha} \equiv V_{\alpha\alpha}$, and the 
expressions for $\cos(K_{1,2} a)$ are given by (\ref{comb}).  So the 
condition $|\cos({\cal K}^{(1,2)} a)| \le 1$ serves for the criterion of 
searching the allowed zones. It is shown in Fig.6 that  the quasi-momenta 
${\cal K}^{(1,2)}$ are close to ones for the uncoupled channels (for $W=1$).

Another peculiarity is that we can {\it create a gap} in the given allowed 
spectral zone  constructing a special multichannel periodic potential. 
In fact, let us consider at first wave motion on a finite interval (period) 
$[0,a]$ with homogeneous boundary conditions. Further, we transform a 
constant interaction matrix $\stackrel\circ{V}_{\alpha \beta }$ defined on   
this interval to $V_{\alpha \beta }(x)$ so that a chosen bound state at 
the energy $E_{n}$ related to $\stackrel\circ{V}_{\alpha \beta }(x)$ is 
raked up to the right boundary by {\it scalar} decrease in the SWV 
$C_{\alpha n}$.  Next, let us continue the resulting matrix block to the 
whole axis with the step $a$ $$V_{\alpha \beta 
}^{per}(x+la)=V_{\alpha \beta }(x), \enskip l=0, \pm 1, \pm 2 ..., \enskip 
0\le x \le a.$$  For this periodic potential, Bloch's solution at the 
energy $E_{n}$ can be constructed in each $l$th interval $[l a, (l+1) a]$ 
from the solutions obtained via the $la$-transfer of $\Psi_{\alpha 
}(x,E_{n})$  over the coordinate line:  $\Psi_{\alpha 
}^{(l)}(x+la,E_{n})=\Psi_{\alpha } (x,E_{n}),\enskip 0\le x \le a$. We have 
the following equalities:  
$\Psi_{\alpha}(0,E_{n})=\Psi_{\alpha}(a,E_{n})=0$ and
$\Psi'_{1}(0,E_{n})/\Psi'_{2}(0,E_{n})=\Psi'_{1}(a,E_{n})/
\Psi'_{2}(a,E_{n})$.
This guarantees the possibility of smooth continuation of the wave
function under the construction $\Psi_{\alpha }^{per}(x,E_{n})$ ($-\infty 
\le x \le \infty$) at the points $l a$ provided that  the module of the 
solution $\Psi_{\alpha }^{(l+1)}(x,E_{n})$ for the $(l+1)$th subsequent 
sector is each time multiplied by a {\it scalar} (i.e., independent of 
$\alpha $) factor $\Theta \equiv \Psi'_{\alpha }(a,E_{n})/\Psi'_{\alpha } 
(0,E_{n})>1 \enskip (\alpha=1,2)$  $$\Psi_{\alpha }^{per}(x+la,E_{n}) = 
(-1)^{l}\Theta ^{l}\Psi_{\alpha }(x,E_{n}), \enskip l=0, \pm 1, \pm 2 ..., 
\enskip 0\le x \le a.$$ Eventually, this leads to an exponential swinging 
divergence  of $\Psi_{\alpha }^{per}(x,E_{n})$ as $x \to  \infty $ at the 
energy $E_{n}$.  It means that this energy belongs to a forbidden zone 
created with the above procedure.

We can transform the periodical multichannel system by creating bound states
inside the allowed or forbidden zones as,  e.g., was shown in \cite{ChZr}.
It is also interesting that using the SUSYQ transformations of 
(\ref{vcomb}) we have an  additional {\it flip effect} of all the 
delta-peaks, as in the one-channel case \cite{ChZr}. 
In fact, after the single SUSYQ transformation (see, e.g. \cite{CZcc}) 
the potential matrix has the form  
\begin{equation}
\hat V(x) = \hat V_{0}(x) - 2 \frac{d}{dx}
[{\hat \Psi }_{0}'(x) {\hat \Psi }_{0}(x)^{-1}]
= -\hat V_{0}(x)+2 {\cal E} + 2 [{\hat \Psi }_{0}'(x)
{\hat \Psi }_{0}(x)^{-1}]^{2},
\label{flip}
\end{equation}
where $\hat V_{0}(x)$ is given by (\ref{vcomb});
${\hat \Psi }_{0}(x)$ represents the corresponding matrix-valued solution
at the factorization energy ${\cal E}$. The term
$-\hat V_{0}(x)$ with the minus sign occurs in the last equality, which
means the flip of all the original $\delta$-peaks which cannot be compensated
for all $x$ by other finite terms.

\section{Conclusion
}

The quintessence of the quantum mechanics is the special relationship of
observables ${\bf S}$ and interactions ${\bf V}$ for wave systems. The 
universal  simple rules for ${\bf V}$ elementary transformations 
caused by elementary ${\bf S}$ variations  in the 
one-channel case were established earlier \cite{ChZ,Z}.

For a long time the multichannel theory has been a 'black box' connecting
input and output data through cumbersome computer calculations. 
The examples presented  in this paper help one to look into 
intimate  physical details of different new effects inherent in coupled 
Schr\"odinger equations.
                        
The considered elements of the multichannel quantum design give some 
notion about elementary  transformations from their complete set for 
systems with the complicated inner structure. No doubt, the
multichannel theory must be richer and more wonderful than in the 
one-channel case.

From a historical point of view, it can be worth  mentioning
that the IP for  multichannel systems with the same thresholds was
considered in the book by Agranovich and Marchenko on multichannel IP 
(see references in \cite{ChS}).
See also the interesting papers \cite{CI,BS,AC,B,N,F} and papers by Cox
(see references in \cite{ChS}).

\section{Acknowledgments
}
The work of V.M.Ch. was supported in part
by INTAS under Grant N96-0457 within the research program of the
International Center for Fundamental Physics in Moscow (ICFPM).
I.V.A. is grateful to Russian Foundation for Basic Research (RFBR) for  
Grant 99-01-01101.

\newpage

FIGURE CAPTIONS

\vspace*{.4cm}
Fig.1.
Two uncoupled channels with the initial infinite rectangular potential wells 
$V_{\alpha \alpha }(x)$ transformed in the following way.
The lowest bound state in the first channel is shifted over the coordinate
axis $x$. The lowest energy level in the second channel is shifted
 over the $E$ axis.
The ground state of the first channel spectral branch was shifted to the
right. All the energy levels in this branch and all other spectral 
weights $C_{n \ne 1}$ remain unperturbed.
To decrease only one spectral weight $\psi '_{1}(x=0)$ of the 
ground state of the first channel, a potential perturbation block 
$\Delta V_{1}(x)$  (barrier + well) is needed.
 The lowest
energy level of the second channel branch was lifted upward without
shifting other levels of the second branch.

\vspace*{.2cm}
Fig.2. A scattering state  is transformed into a bound state at the 
origin in the first uncoupled channel  by increasing its spectral weight 
$C(E_{BSEC})$.  

Soliton-like well in the second channel  serves as a 
carrier for the chosen bound state moving out of the original well when the 
partial channel spectral weight vector component  decreases. 

\vspace*{.2cm}
Fig.3. Increasing first channel parameter ($M_{11}=10^7
\stackrel{\circ}{M}_{11}, 
M_{21}=\stackrel{\circ}{M}_{21}$) for the ground state. 
Both the initial wave functions $\stackrel{\circ}{\Psi }_{1}(x,E_{1}) $ and
$\stackrel{\circ}{\Psi }_{2}(x,E_{1}) $ are concentrated after  the
transformation  in the first channel wave component $\Psi _{1}(x,E_{1}) $ 
and pooled  to the right inside the 
separate soliton-like potential well  $V_{11}(x)$.  The 
initial potential matrix is defined on the half-line and has the form 
$\stackrel{\circ}{V}_{11}(x) = \stackrel{\circ}{V}_{22}(x) = -5, \enskip
\stackrel{\circ}{V}_{12}(x) = \stackrel{\circ}{V}_{21}(x)= 0.3, 
\enskip 0 < x < \pi, \enskip V_{\alpha\beta}(x)=0, \enskip x > \pi$,
the thresholds are $\epsilon_{1}=0, \epsilon_{2}=1$.
The remaining part  of $\Psi _{2}(x,E_{1}) $ in the second channel is so
small that is not visible.

\vspace*{.2cm}
Fig.4. Two degenerated bound states created at $E_{b}=-0.5$ become 'more' 
linearly dependent. This results in splitting of a) the interaction 
matrix and b) wave functions into two parts. The initial system corresponds 
to free motion along the whole line.  a) Two blocks of $V_{\alpha \beta 
}(x)$ for almost equal SWV:  $M_{1}=M_{2}=1;\,\,\,M'_{1}=1,\, M'_{2}=1.01$.  
 The left part of $V_{\alpha \beta }(x)$, a soliton-like well 'moves' to 
$ -\infty $ in the limit  $M'_{2} \to M_{2}$ as is schematically 
shown by dashed lines. b) This well carries also the left part of wave 
 functions to $-\infty $.

\vspace*{.2cm}
Fig.5. a) A potential block (left part of the interaction matrix) without a 
bound state is separated and shifted to the left when two closely spaced 
bound states are created :  $E_{b}=0.5;\,\,\, M_{1}=0,\, M_{2}=1$ 
$E_{b'}=0.501; \,\,\, M'_{1}=1,\, M'_{2}=0.1$.  This block 'moves' to $ 
-\infty $ in the limit $E_{b} \to E_{b'}$ as is schematically shown by 
dashed lines. b) Wave functions remain almost unchanged. They don't "live"
inside this potential block, which is almost transparent if considered 
separately of other part of potential matrix, see \cite{bal} and Fig.8 
there.

\vspace*{.2cm}
Fig.6. a) The functions $\cos({\cal K}^{(1,2)}\pi )$ (solid line) 
 specifying the location of spectral zones for the two-channel 
system are shown.  Allowed and forbidden zones are determined by  
$|\cos({\cal K}^{(1,2)}\pi )|$ $\le 1$ and 
$|\cos({\cal K}^{(1,2)}\pi )| \ge 1$, respectively.  The 
parameters are $V_{1}=6; V_{2}=5; W=1; a = \pi; \epsilon_{1}=0; 
\epsilon_{2}=1$. In the case of uncoupled channels, the corresponding 
curves $\cos(K_{1,2}\pi )$ for separate channels are shifted from one 
another due to the threshold difference and have crossings shown 
by the dashed lines.  Switching on the coupling transforms the crossings 
(dashed lines) into the quasi-crossings so that there are the upper and the 
lower solid lines for two branches of band spectra.  
b) Solid line intervals are forbidden spectral zones of the coupled 
equations (their allowed zones are outside these intervals). 
The solid and the dashed line intervals together represent the forbidden 
zones for any uncoupled equation.  Only blank intervals between vertical 
dashes correspond to the common allowed zones for both uncoupled equations.

 
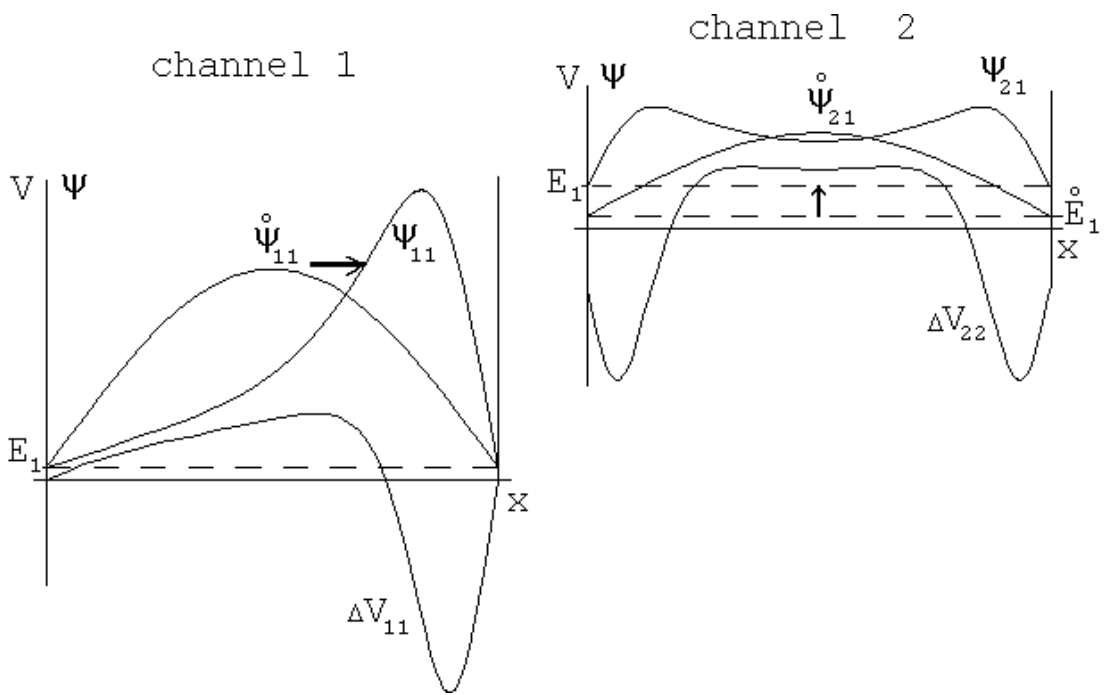
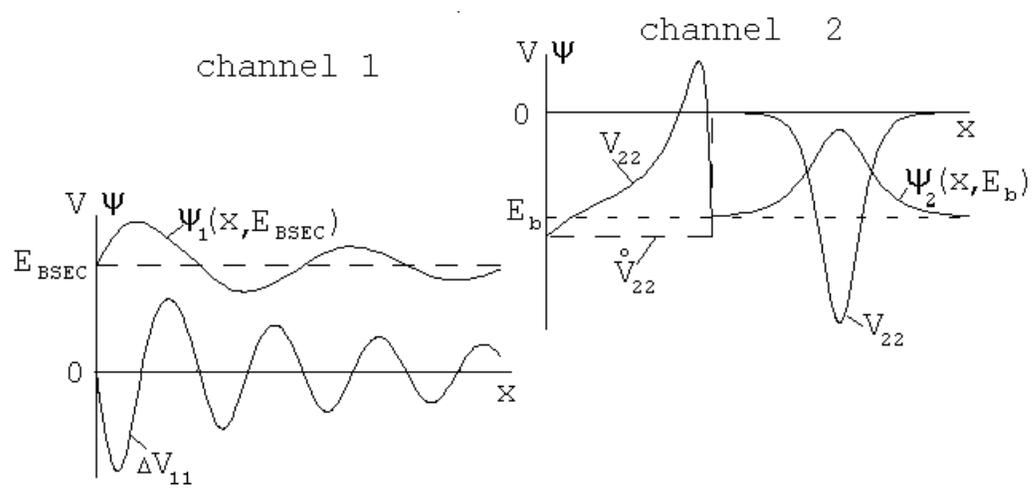
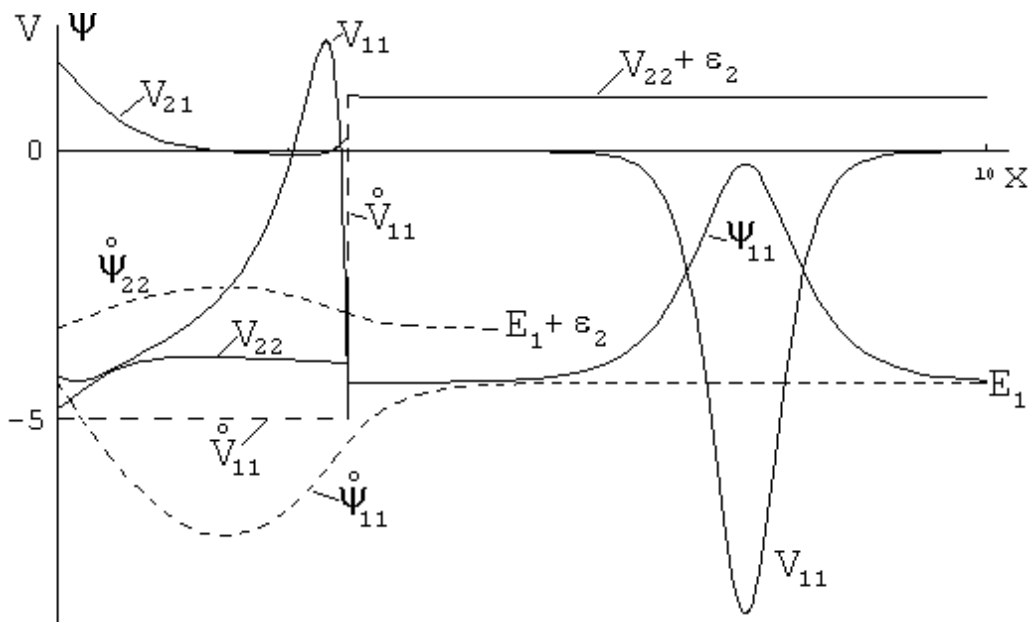
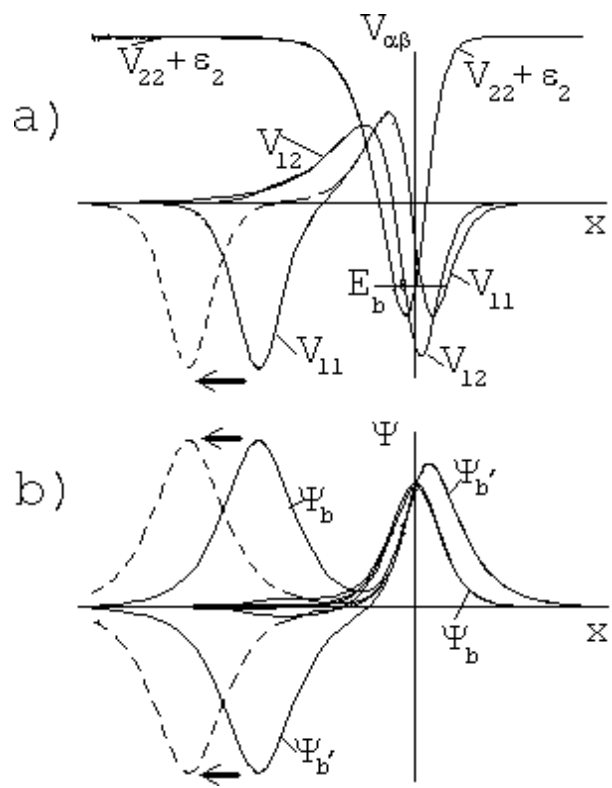
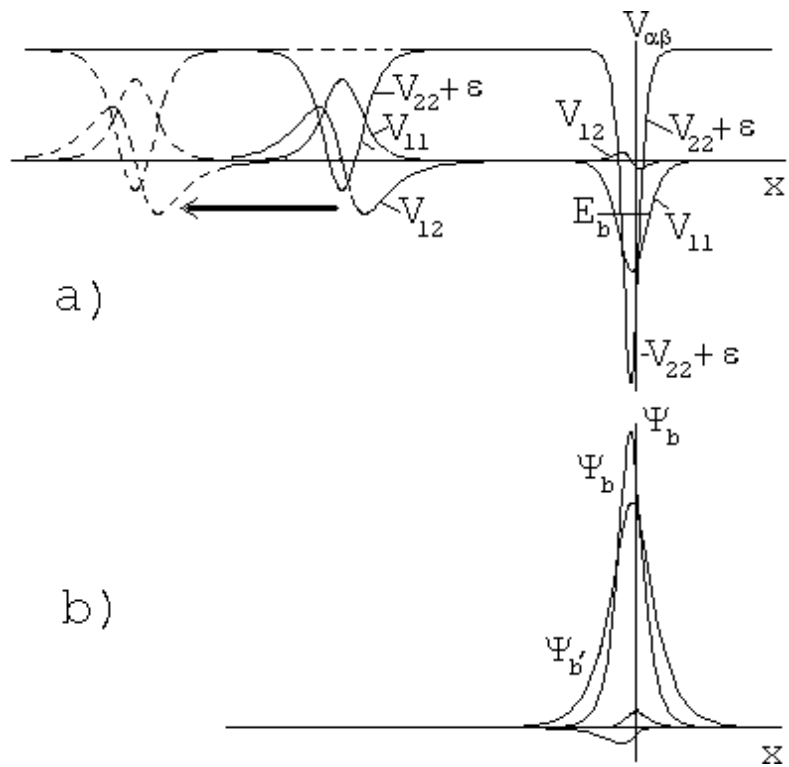
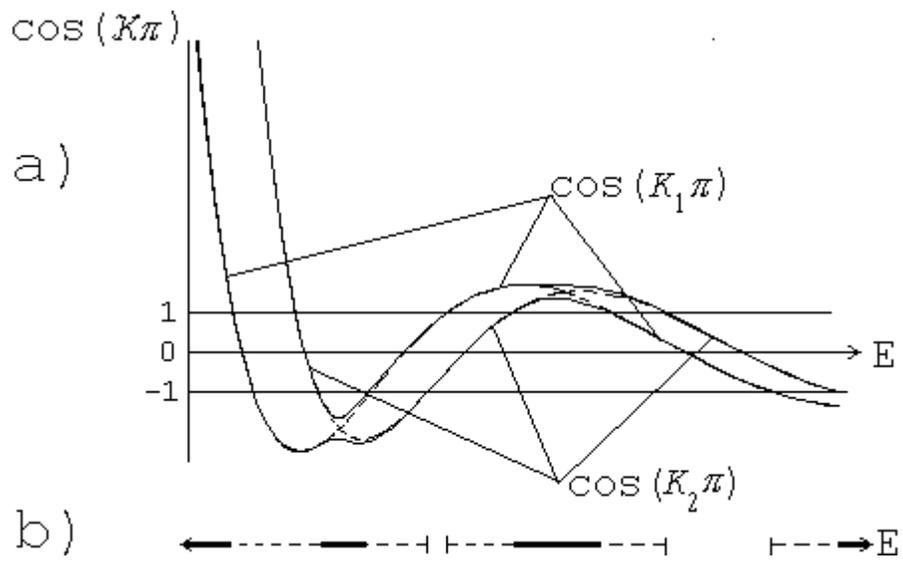
\end{document}